\documentclass[sigconf]{acmart}
\usepackage{subfig}

\AtBeginDocument{%
  \providecommand\BibTeX{{%
    \normalfont B\kern-0.5em{\scshape i\kern-0.25em b}\kern-0.8em\TeX}}}

\copyrightyear{2022}
\acmYear{2022}
\setcopyright{acmlicensed}\acmConference[FDG '22]{FDG '22: Proceedings of the 17th International Conference on the Foundations of Digital Games}{September 5--8, 2022}{Athens, Greece}
\acmBooktitle{FDG '22: Proceedings of the 17th International Conference on the Foundations of Digital Games (FDG '22), September 5--8, 2022, Athens, Greece}
\acmPrice{15.00}
\acmDOI{10.1145/3555858.3555928}
\acmISBN{978-1-4503-9795-7/22/09}
\acmConference['FDG']{Foundations of Digital Games}{September 05--08,
  2022}{Athens, Greece}
\acmISBN{978-1-4503-XXXX-X/18/06}

\begin{document}

\title{SketchBetween: Video-to-Video Synthesis for Sprite Animation via Sketches}

\author{Dagmar Lukka Loftsd{\'o}ttir}
\email{lofts@ualberta.ca}
\affiliation{%
  \institution{University of Alberta}
  \city{Edmonton}
  \state{Alberta}
  \country{Canada}
}

\author{Matthew Guzdial}
\email{guzdial@ualberta.ca}
\affiliation{%
  \institution{University of Alberta}
  \city{Edmonton}
  \state{Alberta}
  \country{Canada}
}

\renewcommand{\shortauthors}{Loftsd{\'o}ttir et al.}

\begin{abstract}
  2D animation is a common factor in game development, used for characters, effects and background art. It involves work that takes both skill and time, but parts of which are repetitive and tedious. Automated animation approaches exist, but are designed without animators in mind. The focus is heavily on real-life video, which follows strict laws of how objects move, and does not account for the stylistic movement often present in 2D animation. We propose a problem formulation that more closely adheres to the standard workflow of animation. We also demonstrate a model, SketchBetween, which learns to map between keyframes and sketched in-betweens to rendered sprite animations. We demonstrate that our problem formulation provides the required information for the task and that our model outperforms an existing method.
  \footnote{The code is available at \url{https://github.com/ribombee/SketchBetween}}
\end{abstract}






\begin{teaserfigure}
  \centering
  \includegraphics[width=\textwidth]{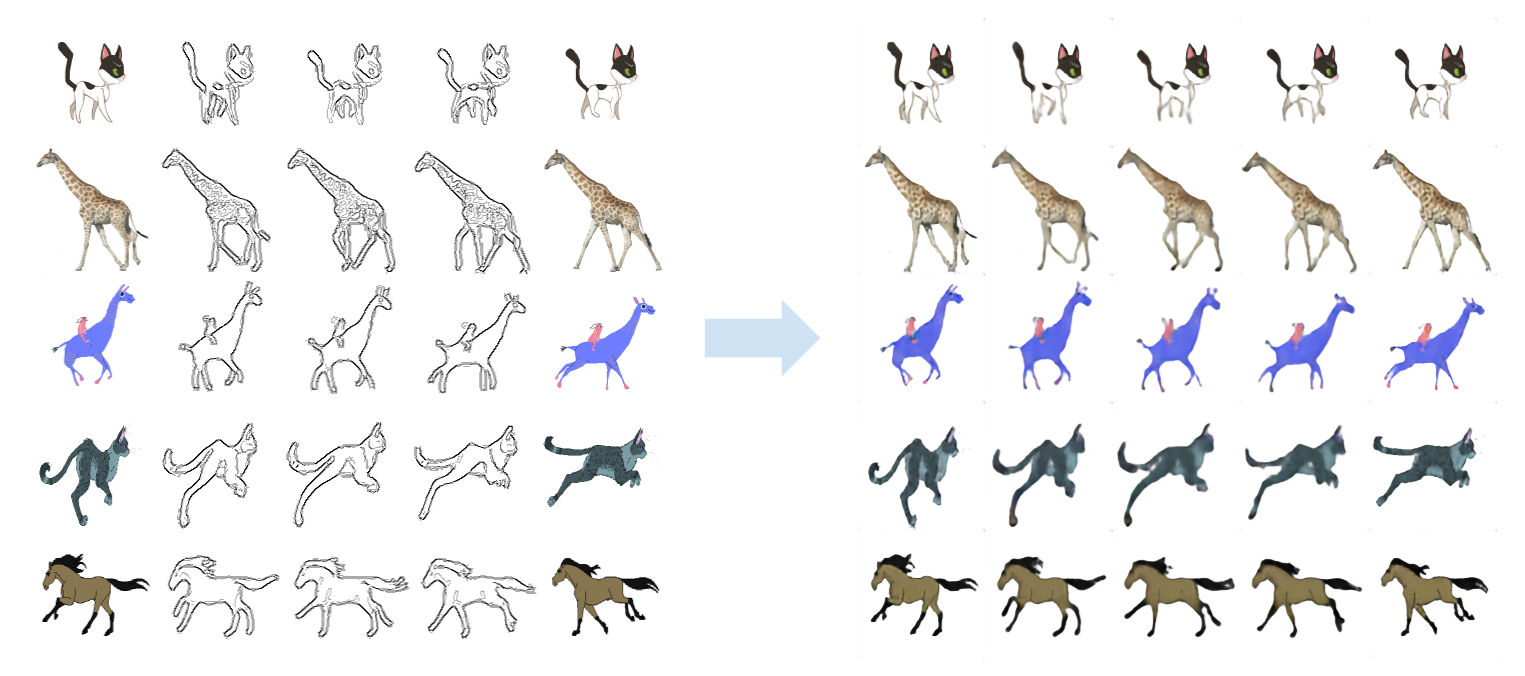}
  \caption{Selected input samples and the corresponding output from SketchBetween.}
  \Description{The input to our SketchBetween model with an arrow pointing to the output.}
  \label{fig:teaser}
\end{teaserfigure}

\maketitle

\section{Introduction}
Games often necessitate the creation of 2D animations for cutscenes, backgrounds, effects or sprites, which are animated art assets of objects or characters in a game. The creation of these types of art assets takes both a high level of skill and a great deal of time \cite{thomas1981illusionchap3}. The process of 2D animation typically breaks down into first creating keyframes that define some action, and then filling in the frames between these keyframes to give the appearance of motion. This is a resource intensive process as it takes someone with a high skill level a long time, but parts of this process are not the most demanding of their skill.

We propose a system which takes as input fully rendered keyframes that define the design and art style of the subject, and sketches for each frame that define the motions in between these keyframes. The system then produces as an output a fully rendered sprite animation.
Existing approaches to video generation and synthesis thus far have not adhered to this standard workflow for animation. Many of the existing methods use previous frames as a prior to generate future frames \cite{DBLP:journals/corr/abs-1711-00937,Walker2021PredictingVW, Rakhimov2021LatentVT}, which requires the full rendering of several adjacent frames. Others rely on transferring a complete motion onto a source image \cite{DBLP:journals/corr/abs-2003-00196}. Both of these methods offer only one perspective of the subject at hand and any information that would be revealed during the motion relies only on what was present in the training set, which may leave a sprite off-model.
We propose a task for the automated rendering of in-between frames given sketches to define the motion, and keyframes to define style. Both a keyframe at the start and at the end are included to capture information about parts of the subject that may be obstructed in either single keyframe. We additionally propose SketchBetween, a system to solve this task. Our system consists of a VQ-VAE \cite{DBLP:journals/corr/abs-1711-00937} that encodes rich information in its latent space about both the desired style and motion and uses this to generate a fully rendered animation. As can be seen in Figure \ref{fig:teaser}, SketchBetween is able to capture stylized motion while staying accurate to character designs. 

There is no direct comparison we can make to other work, as our proposed task is not one that has been studied before. However, it builds upon work in adjacent fields and similar tasks. We evaluate SketchBetween against a strong baseline method adapted to our task using the structural similarity index metric (SSIM) \cite{1284395} and the peak signal-to-noise ratio (PSNR) measured between the original animation and the model's recreation. Compared to the baseline, we see a clear indication of its appropriateness for the target task.

\section{Background}

\subsection{Video Frame Interpolation}
Video Frame Interpolation is a task with the goal of predicting a frame between two adjacent frames in a video.
Machine-learning based methods have shown success on this task, benchmarked on real-life video \cite{DBLP:journals/corr/abs-1810-08768, Li2021VideoFI}.
Most relevant to our work, Siyao Li et al. \cite{DBLP:journals/corr/abs-2104-02495} introduce the problem of video interpolation on animated video. Their method uses segment-guided matching and a recurrent flow refinement network to generate the interpolated frame. This task, however, differs from ours since these methods only produce a single frame at a time via interpolation whereas we transform frames from sketches to fully rendered frames.

\subsection{Video Generation}
Video generation involves a model which predicts future frames given a set of prior frames. There are many approaches in prior work to achieve this goal. 
VQ-VAEs \cite{DBLP:journals/corr/abs-1711-00937} have shown success at this task from their conception, as the original paper demonstrates their ability to generate future frames conditioned on prior frames as well as an action. Expanding upon this work, a number of researchers have used transformers to similarly predict future frames in the latent space of a VQ-VAE \cite{Walker2021PredictingVW, Rakhimov2021LatentVT}.  This system would not fit into an animation workflow, due to the effort required to render several frames at the start of an animation, and because of the loss of control over the motion itself.

\subsection{Video-to-Video Synthesis}
The field of video-to-video synthesis is concerned with mapping one video domain to another \cite{Wang2018VideotoVideoS}. An example of this would be mapping semantic segmentation masks to fully rendered video or mapping video of an action performed by one human onto video of another human. Our task can be considered a subtask of video-to-video synthesis as both the input and output are videos. However, our task relates specifically to animation and not to the more generalized video-to-video synthesis of prior work. Generative adversarial networks (GANs) and methods built on them have previously been applied with good success to several tasks in this field \cite{Wang2018VideotoVideoS, Wang2019FewshotVS, Mallya2020WorldConsistentVS}. GAN-based models also cannot be applied to new subjects without fine-tuning to the new subject. In our case, we want to generate a specific output that's strongly conditioned on an input sketch. As such, a GAN would be less suited to this task than more freeform animation. However, the structure of a VQ-VAE should prove well-suited to this task without requiring a second discriminator model.

\subsection{Image Animation}
Image animation refers to the task of transferring the motion from a video onto an image. The result of this is the subject of the image moving in the same way as the subject of the video. Yoon et al. \cite{PoseGuidedHumanAnimationFromSingleImage} present a method of animating humans from a single image guided by body poses. Their method puts an emphasis on preserving the identity of textures and garments in the synthesized images. Siarohin et al. \cite{DBLP:journals/corr/abs-2003-00196} demonstrate a method which transfers motion from a driving video onto a source image, using keypoint detection and local affine transformations, which are transformations that preserve lines and points locally. The keypoints and their transformations provide additional context to the generator model that produces the frames. However, sprite animation often involves stylized motion that cannot be described with the same methods that perform well with real-life video. An artist may choose to exaggerate shapes and certain aspects of movement to achieve their desired effect to imbue the animation with feelings of weight, speed, or emotion \cite{thomas1981illusionchap3}. We found in our experiments that the work of  Siarohin et al. performed less well on these stylized animations as we demonstrate in our results below. 

\subsection{ML for art generation and representation}
Outside of these specific tasks in the earlier subsections, there are other approaches to generate and represent images related to this work. Both the discrete VAE based model DALL-E \cite{DALLE} and the diffusion-based model GLIDE \cite{GLIDE} have shown a strong ability to generate different art styles through text-to-image generation, however they can't be sufficiently conditioned on temporal information such as adjacent frames. Additionally, VQ-VAEs have shown promise in encoding sprite art, specifically pixel art \cite{AKASH}.

\section{System Overview}
In this section we overview the process of training and applying our SketchBetween system. SketchBetween consists of a VQ-VAE which takes as input the two keyframes and the sketches between them and outputs a fully rendered animation. We train this system using the MGIF dataset \cite{Siarohin2019AnimatingAO} consisting of animations of cartoon animals in motion. We chose five frames for our experiments, but experiments performed with three frames had comparable results. We implemented and trained the system using Keras \cite{chollet2015keras}.

\subsection{Data}
To train our model we require a dataset of existing animations. Therefore we make use of the MGIF dataset \cite{Siarohin2019AnimatingAO}. The MGIF dataset consists of $1000$ videos of cartoon animals walking, running and jumping. Each video has been resized to $128\times128$ pixels and has a white background. There is a good deal of variance in terms of species and art styles present in the data. We use the included train-test split, where $900$ gif format videos are for training and another $100$ are reserved for testing. The videos are of different lengths, and we had to exclude any that were shorter than 5 frames since we chose $N=5$ for our experiments. This exclusion criteria applied to $33$ videos in the training set and $2$ in the test set.

Given our problem formulation in which we transform from keyframes and sketch in-between frames to a final animation we needed to reprocess the data. A dataset including hand-drawn sketches of animations as well as the final output would be ideal. Datasets of still images and corresponding artist sketches exist in the literature \cite{LIPS2019, 7780462, 10.1145/3450626.3459819}. However, no such dataset exists for animations. To generate the input to our model we sample $N$ frames of an animation and generate sketches of all but the first and last frames using Canny edge detection \cite{4767851}, inspired by the method used to train the pix2pix image-to-image translation method \citep{https://doi.org/10.48550/arxiv.1611.07004}. In order to be able to generate convincing sketches for the many different art styles present in the dataset, we opted to average the Canny edges detected for four different kernel sizes ($3$, $5$, $7$ and $9$). The generated input can be seen on the left-hand side of Figure \ref{fig:teaser}. The task is to generate the full $N$ frames, and as such we use the original frames not processed into sketches as our expected output during training.
To achieve better generalization during training, we augment the data by randomly shifting the hue and saturation as well as flipping horizontally before the sketch generation step. Each of these augmentations was applied with a 50\% chance. The hue was shifted randomly by up to 180° and the saturation was randomly increased or decreased by up to 20\%.

\subsection{Model}
\begin{figure*}
    \centering
    \includegraphics[width=\textwidth]{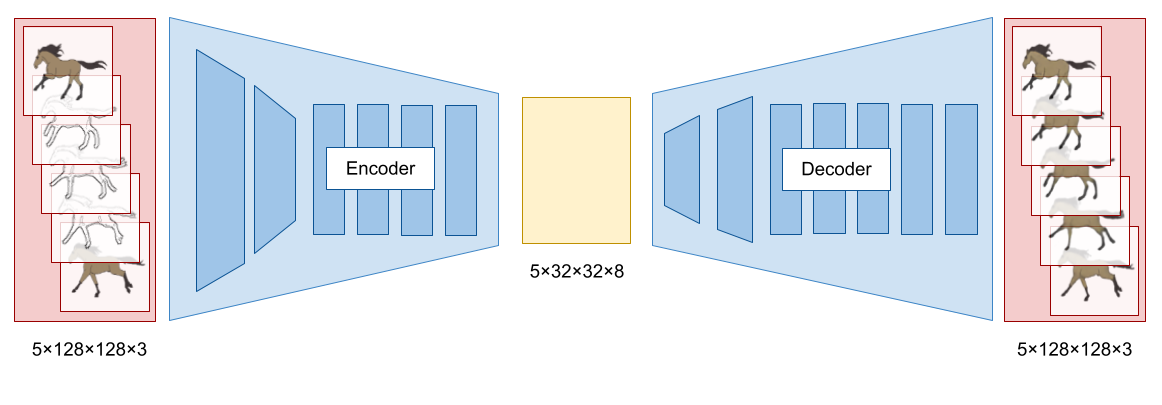}
    \caption{Diagram of the structure of the SketchBetween model. }
    \label{fig:model_diagram}
\end{figure*} 
Our problem formulation requires a model which can take in $N$ frames where frame $0$ and frame $N-1$ are fully rendered and frames $1$ to $N-2$ are sketches. This model must then output a fully rendered animation. 
We chose to train a vector-quantized variational auto-encoder (VQ-VAE) \citep{DBLP:journals/corr/abs-1711-00937} due to their success as a generative model, particularly in the task of art generation and representation \cite{DALLE, AKASH}. 
The inputs to our VQ-VAE are the stacked input frames in RGB format consisting of the keyframes and the sketches between them. The size of our input, barring the batch-size dimension, is $5\times128\times128\times3$. We chose to use RGB because this is the default format of the MGIF dataset. However, this method should work just as well, if not better, if the images are transformed into the HSV format.
We expect this due to the previous success of the HSV format applied to sprite-generating procedural content generation (PCG) tasks \cite{ADRIAN}.
 
An overview of the model structure can be seen in Figure \ref{fig:model_diagram}. The encoder consists of six layers of convolutions with ReLu \cite{ReLU} activation functions and batch normalization \cite{BatchNorm} between each layer of convolutions. We use mostly $3\times3\times3$ kernels, analogous to the encoder structure of GLIDE \cite{DALLE}, throughout our encoder. The exception to this structure are layers of $1\times1\times1$ convolutions in the final two layers to adjust the size of each encoding vector, much like in PixelVQ-VAE \cite{AKASH}. To increase the size of the patch each encoding vector represents, we use a stride of 2 along the x and y axes of each frame in the first two layers. We found that the performance of our model was sensitive to the dimensionality of our encodings (D) and the size of our codebook (C), and our final values for these hyperparameters were D=8 and C=256. The progression of the filter sizes is $3\rightarrow32\rightarrow64\rightarrow64\rightarrow128\rightarrow64\rightarrow D$.
The decoder consists of transposed convolutions with ReLu and batch normalization to project the encoding vectors to the output images. The structure is similar to the encoder, with $3\times3\times3$ convolutions in most layers. The second-to-last layer has $1\times1\times1$ convolutions to decrease the filter sizes towards the end goal of having only 3 filters for the red, green and blue channels of the output images. However, we found empirically that the final decoder layer performed better with a $3\times3\times3$ kernel size. The progression of the filter sizes for the decoder is $D\rightarrow128\rightarrow64\rightarrow64\rightarrow64\rightarrow32\rightarrow16\rightarrow3$. We additionally found that we achieved better performance with a slightly larger decoder than encoder. We anticipate this is because the task of generating embedding representation from the animation is in some way an easier task than generating the fully rendered animation from the embedding representation.

\subsection{Training}
The model was implemented and trained using Keras \cite{chollet2015keras}.
To train SketchBetween we use a loss based on the structural similarity index metric (SSIM) \cite{1284395}, defined as $1-SSIM$, which produced better recreations than a mean-squared-error (MSE) loss. We optimize with a lookahead \cite{Zhang2019LookaheadOK} adam optimizer \cite{Kingma2015AdamAM}. The model is trained for 100 epochs, with a learning rate of 0.001.

The empirical analysis used to determine certain parameters and parts of the model structure were performed on the training set SSIM loss.

\section{Evaluation}

We created SketchBetween as part of a push to make 2D animation more accessible in games. Therefore, human evaluation would be ideal, but as an approximation we provide two forms of quantitative evaluation of our proposed contributions. To evaluate SketchBetween we construct a baseline from a method developed for a related task. To evaluate our proposed problem formulation we perform an ablation study to validate the efficacy of our chosen formulation for the task. 

To evaluate our model on the test set we take every five neighboring frames, including all overlapping sections, and generate inputs for our model. The numbers presented are the reconstruction metrics of only the three in-between frames, excluding the first and last frame as their reconstruction is less relevant to the task. The metrics improve marginally if the two keyframes' reconstructions are included, but we felt the relevant component of our model was the reconstruction of the sketched in-between frames. We use two metrics to evaluate the quality of reconstructions. The structural similarity index metric (SSIM) \cite{1284395} as well as the peak signal-to-noise ratio (PSNR) between the original frame and the recreation. The SSIM measures the structural similarity between the original and the reconstruction in terms of luminance, contrast and structure. The PSNR measures the ratio between the maximum values of a signal and the noise present.

\subsection{Baseline}
We evaluate our method against a baseline that we have constructed by adapting a first-order motion model (FOMM) \cite{DBLP:journals/corr/abs-2003-00196} to our task. FOMM transfers motion from a video onto an image. It is only trained on the videos which drive the motion, and makes the assumption that the image derives from the same distribution. To adapt it to our task we train the FOMM on a dataset consisting of the original MGIF dataset with the addition of sketched versions of each gif using our sketch generation method. This ensures that keypoint detection is trained on both sketches and rendered images simultaneously and should provide a model which can transfer sketch motion onto a rendered keyframe. The baseline is then used to transfer the motion from the generated sketches of the test set onto the first fully rendered frame. The SSIM and PSNR measured are averaged over every frame in the generated animation. FOMM generates each animation in full from the sketched version and the first frame so there is a difference in the number of datapoints being evaluated as SketchBetween generates 3 frames from sketches for every 5 adjacent frames.

\subsection{Ablation Study}
We perform an ablation study to show the efficacy of the key components of our problem formulation. Ideally we would use human evaluation, but as an approximation we perform an ablation study and evaluate the reconstruction metrics of the ablated models with and without certain components. Firstly, we compare our system with and without the sketches between the keyframes. To do this we train a SketchBetween model on our dataset without generating sketches for the in-between frames, leaving them blank. We did this because we believe the sketches provide valuable information about the motion present in the animation. Secondly we compare our system with and without the second keyframe in place. To do this we train a SketchBetween model on our dataset but generate a sketch for the final frame in place of the fully rendered image. We believe that both the first and last keyframe are of value to the model to properly handle cases where parts of the subject are obstructed at the start or end and become revealed during the motion. In both cases the metrics are reported on a version of the test set which has the same ablation to its structure.

\section{Results}
\begin{table}
\centering
\caption{Comparison between our method and baselines on the MGIF dataset. $\uparrow$ means that a higher value is better.}
\begin{tabular}{lll}
\toprule
Model      & SSIM ($\uparrow$) & PSNR ($\uparrow$) \\
\midrule
FOMM &     0.799  &     19.46 \\
Ours       &   \textbf{0.943}      &     \textbf{27.48}     \\
\bottomrule
\end{tabular}
\label{Tab:baselinetable}
\end{table}

\begin{table}
\centering
\caption{Comparison between our complete method and ablated versions of it. $\uparrow$ means that a higher value is better.}
\begin{tabular}{lll}
\toprule
Model      & SSIM ($\uparrow$) & PSNR ($\uparrow$) \\
\midrule
No sketch &    0.883     &    20.19     \\
No final image &     0.938    &   25.4      \\
Full model       &   \textbf{0.943}      &     \textbf{27.48}   \\  
\bottomrule
\end{tabular}
\label{Tab:ablationtable}
\end{table}

\begin{figure}
    \centering
    \includegraphics[width=0.09\textwidth]{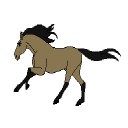}
    \includegraphics[width=0.09\textwidth]{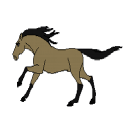}
    \includegraphics[width=0.09\textwidth]{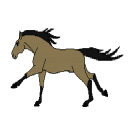}
    \includegraphics[width=0.09\textwidth]{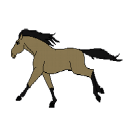}
    \includegraphics[width=0.09\textwidth]{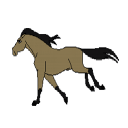}
    
    \includegraphics[width=0.09\textwidth]{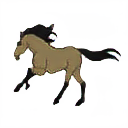}
    \includegraphics[width=0.09\textwidth]{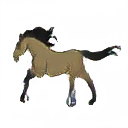}
    \includegraphics[width=0.09\textwidth]{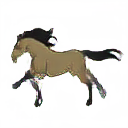}
    \includegraphics[width=0.09\textwidth]{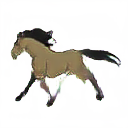}
    \includegraphics[width=0.09\textwidth]{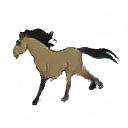}
    
    \includegraphics[width=0.09\textwidth]{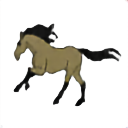}
    \includegraphics[width=0.09\textwidth]{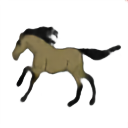}
    \includegraphics[width=0.09\textwidth]{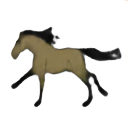}
    \includegraphics[width=0.09\textwidth]{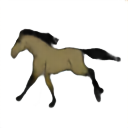}
    \includegraphics[width=0.09\textwidth]{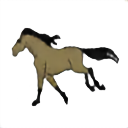}
    
    \caption{Visual comparison of the recreation of a horse running animation. The top row is the original animation. The middle animation is the FOMM recreation and the bottom row is the SketchBetween recreation.}
  \Description{Three rows of images. The top row shows an artist rendered animation of a horse running. The middle row shows a recreation of that animation with the FOMM baseline. The bottom row shows a recreation of that animation with SketchBetween.}

    \label{fig:baselinehorse}
    
\end{figure}

\begin{figure}
    \centering
    \includegraphics[width=0.09\textwidth]{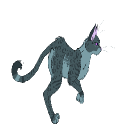}
    \includegraphics[width=0.09\textwidth]{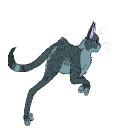}
    \includegraphics[width=0.09\textwidth]{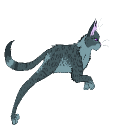}
    \includegraphics[width=0.09\textwidth]{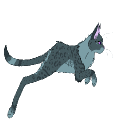}
    \includegraphics[width=0.09\textwidth]{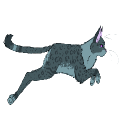}
    
    \includegraphics[width=0.09\textwidth]{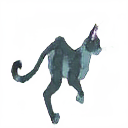}
    \includegraphics[width=0.09\textwidth]{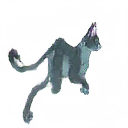}
    \includegraphics[width=0.09\textwidth]{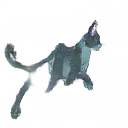}
    \includegraphics[width=0.09\textwidth]{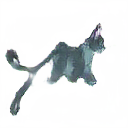}
    \includegraphics[width=0.09\textwidth]{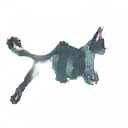}
    
    \includegraphics[width=0.09\textwidth]{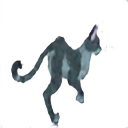}
    \includegraphics[width=0.09\textwidth]{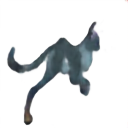}
    \includegraphics[width=0.09\textwidth]{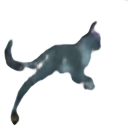}
    \includegraphics[width=0.09\textwidth]{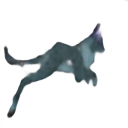}
    \includegraphics[width=0.09\textwidth]{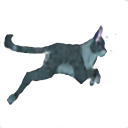}
    
    \caption{Visual comparison of the recreation of a cat leaping animation. The top row is the original animation. The middle animation is the FOMM recreation and the bottom row is the SketchBetween recreation.}
    \label{fig:baselinecat}
  \Description{Three rows of images. The top row shows an artist rendered animation of a cat running. The middle row shows a recreation of that animation with the FOMM baseline. The bottom row shows a recreation of that animation with SketchBetween.}
\end{figure}

\begin{figure}
    \centering
    \includegraphics[width=0.09\textwidth]{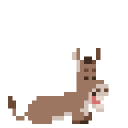}
    \includegraphics[width=0.09\textwidth]{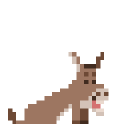}
    \includegraphics[width=0.09\textwidth]{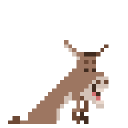}
    \includegraphics[width=0.09\textwidth]{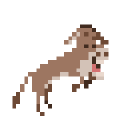}
    \includegraphics[width=0.09\textwidth]{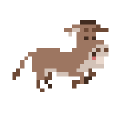}
    
    \includegraphics[width=0.09\textwidth]{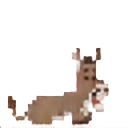}
    \includegraphics[width=0.09\textwidth]{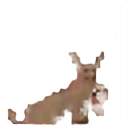}
    \includegraphics[width=0.09\textwidth]{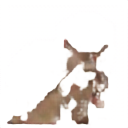}
    \includegraphics[width=0.09\textwidth]{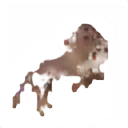}
    \includegraphics[width=0.09\textwidth]{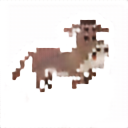}
    
    \caption{Visual comparison of the recreation of a pixel-art donkey jumping animation. The top row is the original animation. The bottom row is the animation generated by SketchBetween.}
      \Description{Three rows of images. The top row shows an artist rendered animation of a pixel-art donkey jumping. The middle row shows a recreation of that animation with the FOMM baseline. The bottom row shows a recreation of that animation with SketchBetween.}
    \label{fig:FailureDonkey}
\end{figure}

A comparison of the results of our model and the baseline can be seen in Table \ref{Tab:baselinetable}. Our method achieves higher scores on this task for both metrics. Additionally, a visual comparison of selected representative recreations can be seen in Figures \ref{fig:baselinehorse} and \ref{fig:baselinecat}. The baseline fails to capture changes in the shapes of objects that are not affine transformations of a part of the subject, but instead animated stylistically, such as the horse's mane and tail flowing behind it. 

The results of our ablation study on the task formulation can be seen in Table \ref{Tab:ablationtable}. We see that the inclusion of all elements does provide the best results according to our metrics. The inclusion of the sketches is especially important. This is in line with the intuition of sketches providing valuable information about how the subject should be rendered in the in-betweens. However, for our dataset, the inclusion of the final keyframe is not hugely significant. This is perhaps because the animations provided in the MGIF dataset have similar motions where obstructed portions of the subject are most frequently limbs, which are typically symmetrical.

\section{Case Study}
Since SketchBetween is intended to be used as a tool by animators we performed a case study of how it might work on real-life examples, such as walk cycles for game characters. We obtained $4$ animations from $3$ different artists, where the artists rendered the first and last frames and sketched three frames of motion in between. The artists were asked to provide a short animation of an animal in motion. Artwork was provided by Twitter users @TeethyFish (fish, chinchilla) and @K3rryberry (bunny) and the first author of this paper (fox).

\begin{figure}
    \centering
    \includegraphics[width=0.09\textwidth]{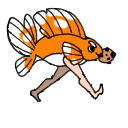}
    \includegraphics[width=0.09\textwidth]{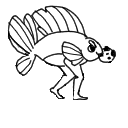}
    \includegraphics[width=0.09\textwidth]{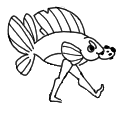}
    \includegraphics[width=0.09\textwidth]{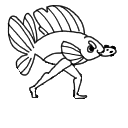}
    \includegraphics[width=0.09\textwidth]{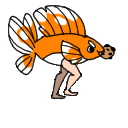}
    
    \includegraphics[width=0.09\textwidth]{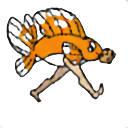}
    \includegraphics[width=0.09\textwidth]{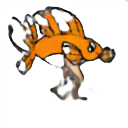}
    \includegraphics[width=0.09\textwidth]{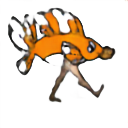}
    \includegraphics[width=0.09\textwidth]{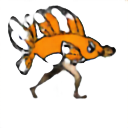}
    \includegraphics[width=0.09\textwidth]{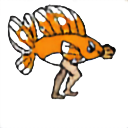}
    
    \hrule
    
    \includegraphics[width=0.09\textwidth]{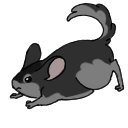}
    \includegraphics[width=0.09\textwidth]{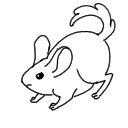}
    \includegraphics[width=0.09\textwidth]{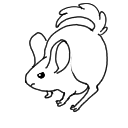}
    \includegraphics[width=0.09\textwidth]{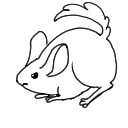}
    \includegraphics[width=0.09\textwidth]{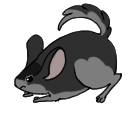}
    
    \includegraphics[width=0.09\textwidth]{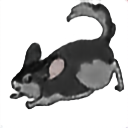}
    \includegraphics[width=0.09\textwidth]{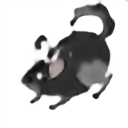}
    \includegraphics[width=0.09\textwidth]{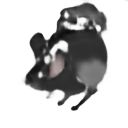}
    \includegraphics[width=0.09\textwidth]{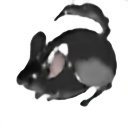}
    \includegraphics[width=0.09\textwidth]{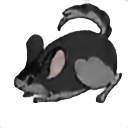}
    
    \hrule
    
    \includegraphics[width=0.09\textwidth]{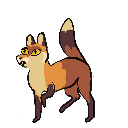}
    \includegraphics[width=0.09\textwidth]{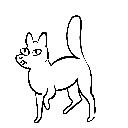}
    \includegraphics[width=0.09\textwidth]{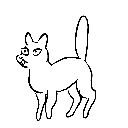}
    \includegraphics[width=0.09\textwidth]{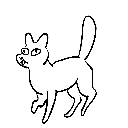}
    \includegraphics[width=0.09\textwidth]{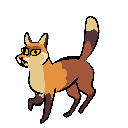}
    
    \includegraphics[width=0.09\textwidth]{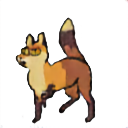}
    \includegraphics[width=0.09\textwidth]{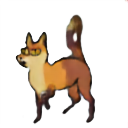}
    \includegraphics[width=0.09\textwidth]{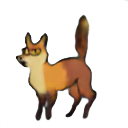}
    \includegraphics[width=0.09\textwidth]{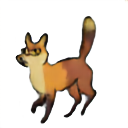}
    \includegraphics[width=0.09\textwidth]{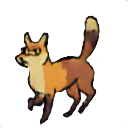}
    
    \hrule
    
    \includegraphics[width=0.09\textwidth]{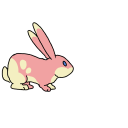}
    \includegraphics[width=0.09\textwidth]{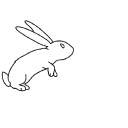}
    \includegraphics[width=0.09\textwidth]{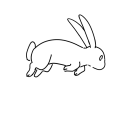}
    \includegraphics[width=0.09\textwidth]{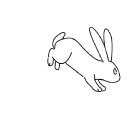}
    \includegraphics[width=0.09\textwidth]{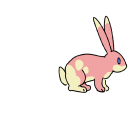}
    
    \includegraphics[width=0.09\textwidth]{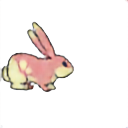}
    \includegraphics[width=0.09\textwidth]{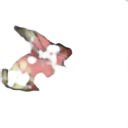}
    \includegraphics[width=0.09\textwidth]{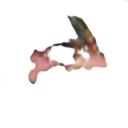}
    \includegraphics[width=0.09\textwidth]{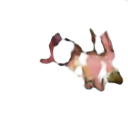}
    \includegraphics[width=0.09\textwidth]{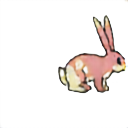}
    
    \caption{Animations by various artists. For each section the top row is an animation with artist-drawn sketches between the first and last frame. The bottom row is the SketchBetween output for this animation.}
    \label{fig:CaseStudy}
      \Description{Four subfigures which each have two rows of images. Each subfigure's top row of images is a hand-drawn animation consisting of two keyframes that are fully rendered with three sketches between. The bottom row is the output of SketchBetween given the top row as input. The first subfigure is of a koifish with a cookie in its mouth and human legs walking. The second is of a chinchilla running. The third of a fox stomping, and the fourth of a bunny leaping.}
\end{figure}

We ran these animations through SketchBetween to see how the model handles human-made sketches. The results of this can be seen in Figure \ref{fig:CaseStudy}. Notably some patterns on the animals that would have been added to the sketches via the canny edge detection are not present in the human-made sketch. The model still attempts to generate them, but they are less consistent and crisp than patterns seen in Figure \ref{fig:teaser}. Prior to running SketchBetween on these animations we were convinced that it would work best on the chinchilla and fox, and worst on the fish. We were surprised to see that the model performed best on the fish and the fox and worst on the chinchilla. We speculate that the large areas with low texture on the chinchilla may have contributed to the amount of artifacts in the output. It should be noted that while the model performs well on the fox, this was drawn by one of the authors and thus they could have unconsciously drawn a well-suited example to SketchBetween. Notably this example includes smaller motion than in the other examples. Conversely, the bunny example is of a creature jumping like in Figure \ref{fig:FailureDonkey} which depicts a large motion at a low framerate which the model seems to handle poorly.

The artists have visibly different sketching styles but notably all the human-made sketches are fairly clean and do not contain much of the artifacts that artists may use to assist in structuring the form of their artwork. It is therefore unclear to what extent SketchBetween can handle the variance of sketching methods used by different artists.
We leave a full investigation of this topic to future work.

\section{Discussion and Future Work}

We have shown through our experiments that our model outperforms a baseline for our task, and that the formulation of this task provides sufficient information to construct models with good outcomes. The system may therefore be able to save artists' time when producing animations. It is difficult to quantify how much time can be saved by using a tool such as this due to the high variance in how artists operate, resulting from differences in process and level of skill. However, the proposed tool shaves off one of the more tedious parts of the animation process.

Our model does have certain shortcomings. There are a number of videos in the test set for which our model performs quite poorly. These are videos with motions that are not common in the training set, one example can be seen in Figure \ref{fig:FailureDonkey}. Jumping is an action that is not common in the dataset and the particular stylization of this donkey's jump makes it an outlier. This effect can also be seen in the bunny in Figure \ref{fig:CaseStudy}.

Additionally, our output videos are more blurry than desired, and details tend to be lost. A future avenue of research could be to explore diffusion-based video-to-video synthesis, since our method is analogous to a sort of inpainting of the in-between frames, and diffusion models have shown success at inpainting \cite{GLIDE}. Another potential direction is to apply an adversarial loss \cite{Wang2018VideotoVideoS, Wang2019FewshotVS, Mallya2020WorldConsistentVS}, or to use a more refined loss than our SSIM-based one. For example, one based on analogs between frames, inspired by previous work in deep visual analogy-making \cite{NIPS2015_e0741335}. 

The case study exposed further avenues for improvement. It highlights a weakness of the dataset used to train the model, which is that it consists of only three types of movements: walking, running and jumping. The dataset also largely contains animations that are produced at a high frames-per-second (fps) and therefore only have subtle changes between adjacent frames. The effect of this is that animations with more drastic changes, such as the bunny in Figure \ref{fig:CaseStudy} are poorly handled. This could be mitigated with more data, but also by omitting some number of frames during training to artificially decrease the fps. The case study also suggests that adding more variance into the generated sketches to capture more art styles would be valuable. Adapting this method to be more iterative could also prove to be useful, as design tools with iterative design processes have been shown to be effective \cite{AniMesh}

The ablation study shows that while the system performs better with the sketches available between the keyframes, there is value in exploring a system which omits this additional work for artists. In animations with a high fps this may shave additional time off, however it may negatively impact animations with large motions between keyframes such as the bunny in Figure \ref{fig:CaseStudy} where there is more motion information in the sketches.

\section{Conclusions}
This paper proposes a problem formulation and model solution for sprite animation generation. We focus on rendering sketches between rendered keyframes. We additionally propose a method to do this using a VQ-VAE called SketchBetween. We showed that our method outperforms a strong baseline, and that the formulation of the task is valuable through an ablation study. Given our results we believe that this research direction has the potential to help democratize 2D animation in game development.

\begin{acks}
We acknowledge the support of the Natural Sciences and Engineering Research Council of Canada (NSERC) and Alberta Machine Intelligence Institute (Amii). 
\end{acks}

\bibliographystyle{ACM-Reference-Format}
\bibliography{SketchBetween}

\end{document}